\documentclass[prl,twocolumn,showpacs,showkeys]{revtex4}

\usepackage{graphicx}
\usepackage{dcolumn}
\usepackage{bm}
\usepackage{amssymb}

\begin{document}
\title{Has the Photon an Anomalous Magnetic Moment?}
\author{S. Villalba-Ch\'avez$^{\dag\ddag}$ and H. P\'erez-Rojas$^{\ddag}$}

\affiliation{$^\dag$ $^1$ Abdus Salam International Centre for
Theoretical Physics, Strada Costiera 11, I-34014, Trieste Italy.\\
$^\ddag$Instituto de Cibernetica, Matematica y Fisica, Calle E
309, Vedado, Ciudad Habana, Cuba}

\date{July, $21 \ \ 2006$}

\begin{abstract}
Due to its interaction with the virtual electron-positron field in
vacuum, the photon exhibits a nonzero anomalous magnetic moment
whenever it has a nonzero transverse  momentum component to an
external constant magnetic field. At low and high frequencies this
anomalous magnetic moment behaves as paramagnetic, and at energies
near the first threshold of pair creation it has a maximum value
greater than twice the electron anomalous magnetic moment. These
results might be interesting in an astrophysical and cosmological
context.
\end{abstract}



\maketitle

It was shown by Schwinger\cite{Schwinger} in 1951 that electrons
get an anomalous magnetic moment $\mu^\prime=\alpha/2\pi\mu_B$
(being $\mu_B=e\hbar/2m_0c$ the Bohr magneton) due to radiative
corrections in quantum electrodynamics (QED), that is, due to the
interaction of the electron with the background virtual photons
and electron-positron pairs. We want to show that also, due to the
interaction with the virtual quanta of vacuum,  an anomalous
photon magnetic moment arises. It is obtained from the expression
for the photon self-energy in a magnetic field, calculated by
Shabad\cite{shabad1,shabad2} in an external constant magnetic
field $\Pi_{\mu\nu}(x,x^{\prime\prime}\vert A^{ext})$ by starting
from the electron-positron Green function in the Furry picture,
and by using the Schwinger proper time method. The expression
obtained was used by Shabad \cite{shabad2} to investigate the
photon dispersion equation in vacuum in presence of an external
magnetic field. It was found a strong deviation from the light
cone curve near the energy thresholds for pair creation, which
suggests that  the photon propagation behavior in the external
classical magnetic field is strongly influenced by the virtual
electron-positron pairs of vacuum near these thresholds, showing a
behavior similar to that of a massive particle. These phenomena
become especially significant near the critical field $B_c=m_0^2/e
\sim 4,41 \cdot 10^{13}$ Gauss, where $m_0, e$ are respectively
the electron mass and charge.

The photon magnetic moment might have astrophysical and
cosmological consequences. For instance, photons passing by a
strongly magnetized star, would experience an additional shift to
the usual gravitational one produced by the star mass.

In presence of an external field the current vector is non
vanishing $j(x)_{\mu}=ie Tr\gamma_{\mu} G(x,x|A^{ext}) \neq 0$,
where $G(x,x^{\prime}|A^{ext})$ is the electron-positron Green's
function in the external field. By calling the total
electromagnetic field by $A^t_{\mu}= A^{ext}_{\mu}+A_{\mu}$, the
QED Schwinger-Dyson equation for  the photon field  $A_\mu(x)$,
propagating in the external field $A_\mu(x)^{ext}$ is
\begin{equation} \left[\square
\eta_{\mu\nu}-\partial_\mu\partial_\nu\right]
A^\nu(x)+\int\Pi_{\mu\nu}(x,x^\prime\vert A^{ext})
A^\nu(x^\prime) d^4x^\prime=0,\label{sdpmBF}
\end{equation}
where $\mu,\nu=1,2,3,4$. The expression  (\ref{sdpmBF}) is
actually the set of Maxwell equations in a neutral polarized
vacuum, where the second term corresponds to the approximation of
the four-current linear in $A_{\mu}$, where the coefficient is the
polarization operator $\delta j_\mu (x)/\delta A^t_\nu
(x^{\prime\prime})|_{A^t=A^{ext}}=\Pi_{\mu\nu}(x,x^{\prime\prime}\vert
A_\mu^{ext})$. The external (constant and homogeneous) classical
magnetic field is described by
$A_\mu^{ext}(x)=1/2F_{\mu\nu}^{ext}x^\nu$, where the
electromagnetic field tensor $F_{\mu \nu}^{ext}=\partial_\mu
A_\nu^{ext}-\partial_\nu A_\mu^{ext}=B (\delta_{\mu 1}\delta_{\nu
2}-\delta_{\mu 2}\delta_{\nu 1})$
 and $F^*_{\mu
\nu}=\frac{i}{2}\epsilon_{\mu \nu \rho \kappa}F^{\rho \kappa}$ is
its dual pseudotensor.

To understand what follows it is necessary to recall some basic
results  developed in refs. \cite{shabad1},\cite{shabad2}. The
presence of the constant magnetic field creates, in addition to
the photon momentum four-vector $C^{4}_\mu=k_\mu$, three other
orthogonal four-vectors which we write as four-dimensional
transverse $k_\mu C^{i\mu}=0$ for $i=1,2,3$. These are $C^{1}_\mu=
k^2 F^2_{\mu \lambda}k^\lambda-k_\mu (kF^2 k)$,
$C^{2}_\mu=F^{*}_{\mu \lambda}k^\lambda$, $C^{3}_\mu=F_{\mu
\lambda}k^\lambda$ ($C^{1,2,3}_{\mu}k^{\mu}=0$). We have
$C^{4}_\mu C^{4\nu}=k_\mu k^\nu=0$ on the light cone.  One gets
from these four-vectors three basic independent scalars $k^2$,
$kF^2k$, $kF^{*2}k$, which in addition to the field invariant
${\cal F}=\frac{1}{4}F_{\mu \rho}F^{\rho \mu}=\frac{1}{2}B^2$, are
a set of four basic scalars of our problem.

In momentum space it can be written the eigenvalue equation
\cite{shabad1}
\begin{equation}
\Pi_{\mu\nu}(k,k^{\prime\prime}\vert A_\mu^{ext})=\sum_i
\pi^{(i)}_{n,n^\prime} a^{(i) \nu }a^{(i)}_\mu/(a^{(i)\nu
}a^{(i)}_\nu ) \label{2}
\end{equation}
In correspondence to each eigenvalue $\pi^{(i)}_{n,n^\prime}$
$i=1,2,3$ there is an eigenvector $a^{(i)\nu }$. The set $a^{(i)\nu }$ is obtained
by simply normalizing the set of four vectors $C^{i}_\mu$.
($C^{4}_\mu=k_\mu$ leads to a vanishing eigenvalue due to the
four-dimensional transversality property
$\Pi_{\mu\nu}(k,k^{\prime\prime}\vert A_\mu^{ext})k_\mu=0$). The
solution of the equation of motion (\ref{sdpmBF}) can be written  as a
superposition of eigenwaves given by
\begin{equation}
A_\mu(k)=\sum_{j=1}^4 \delta(k^2-\pi_j)a_\mu^j(k) \label{3}
\end{equation}
By considering $a^{(i)}_\mu (x)$ as the electromagnetic four
vector describing the eigenmodes, it is easy to obtain the
corresponding electric and magnetic fields of each mode ${\bf
e}^{(i)}= \frac{\partial }{\partial
x_0}\vec{a}^{(i)}-\frac{\partial }{\partial {\bf x}}a^{(i)}_0$,
${\bf h}^{(i)}=\nabla\times\vec{a}^{(i)}$ (see \cite{shabad2}).

From now on we specialize in a frame in which $x_3||B$. Then
$kF^2k/2\mathcal{F}=-k_{\perp}^2$ and we name $z_1= k^2 +
kF^2k/2\mathcal{F}=k_{\parallel}^2-\omega^2$. The previous results
(see \cite{shabad2}) indicate the existence of three dispersion
equations with the following structure
\begin{equation}
k^2=\pi^{(i)}\left(z_1,k_{\perp}^2,eB\right).\
\\ \ i=1,2,3 \label{egg}
\end{equation}
The eigenvalues $\pi^{(i)}$ contain only even functions of the
external field through the scalars  $kF^2k$, $kF^{*2}k$, and $e
\sqrt{2 \cal F}=eB$ and can be expressed as a functional expansion
in series of  even powers of the product $e A_\mu^{ext}$
\cite{Fradkin}.

One can solve (\ref{egg}) for $z_1$
in terms of $k_{\perp}^2$. It results
\begin{equation}
\omega^2=\vert\textbf{k}\vert^2+f_i\left(k_{\perp}^2,B\right)
\label{eg2}
\end{equation}
The term $ f_i$ contains the interaction of the photon with  the
virtual $e^{\pm}$ pairs in the external field in terms of the
variables $k_{\perp}^2,B$. As it is shown in \cite{shabad2}, it
makes the photon dispersion equation to have a drastic departure
from the light cone curve near the energy thresholds for free pair
creation,

We are thus in conditions to  define an anomalous magnetic moment
for the photon as $\mu_\gamma=-\partial \omega/\partial B$. Then
$\mu_\gamma$ is a function of $B$. For weak fields ($B \ll B_c$),
and frequencies small enough (see below), the function $f_i$ can be
written as linear in $B$, the resulting dispersion law being then
\begin{equation}
\omega=\vert\textbf{k}\vert- \mu_{\gamma} B \label{de0}
\end{equation}
The first term corresponds to the light cone equation, which is
modified by the second, which contains the  contribution of the
photon magnetic moment.

 The gauge invariance property
$\pi^{(i)}(0,0)=0$ implies that the function  $f_i(k_{\perp}^2,B)$
vanishes when $k_{\perp}^2=0$ \cite{proceeding}. This means that,
due to gauge invariance, when the propagation is parallel to
$\bf{B}$,   $\mu_{\gamma}$  vanishes. Thus, in every mode of
propagation $\mu_\gamma=0$ if $k_\perp=0$. Therefore the photon
magnetic moment depends essentially on the perpendicular momentum
component and this determines the optical properties of the quantum
vacuum in presence of $B$.

As a result, the problem of the propagation of light in empty space,
in presence of an external magnetic field is similar to the problem
of the dispersion of light in an anisotropic medium, where the role
of the medium is played by the polarized vacuum in the external
magnetic field. An anisotropy is created by the preferred direction
in space along $\textbf{B}$. Therefore, the refraction index
$n^{(i)}=\vert\textbf{k}\vert/\omega_i$ in mode $i$  is given in the
case in which the approximate expression (\ref{de0}) is valid as
\begin{equation}
n^{(i)}=1+\frac{\mu_\gamma B }{\vert\textbf{k}\vert} \label{in}
\end{equation}
For
parallel propagation, $k_\perp=0$, for any mode it is obviously $n_i=1$.

In \cite{shabad1,shabad2} (see also \cite{proceeding}) it was
shown that  $\Pi_{\mu \nu}$ has singularities starting the value
$z_1=-4m_0^2$, which is the first threshold for pair creation,
corresponding to Landau quantum numbers $n=n^{\prime}=0$. Other
pair creation thresholds are given by
$k_{\perp}^{\prime}=m_0^2[(1+2 n B/B_c)^{1/2}+(1+2n^\prime
B/B_c)^{1/2}]^2$, with the electron and positron in excited Landau
levels $n,n^{\prime}\neq 0$). In what follows we will work in  the
transparency region, that is, out from the region for absorption
due to the pair creation i.e., $\omega^2-k_\parallel^2\leq
k_{\perp}^{\prime 2}$ (\textit{i.e.} within the kinematic domain,
where $\pi_{1,2,3}$ are real.  We will be interested in two
limits, i.e., when its energy is near the first pair creation
threshold energy (and the magnetic field $B \sim B_c$), and when
it is much smaller than it,  $4m_0^2\gg \omega^2 $ and  small
fields $(B\ll B_c)$,  in the one loop approximation. Below the
first threshold the eigenvalues corresponding to the first and
third modes do not contribute, whereas the second mode it is shown
in \cite{prd}, by using the formalism developed iny \cite{shabad1,
shabad2} that the eigenvalue near this threshold is in that limit
\begin{equation}
\pi_2=-\frac{2\mu^{\prime}B}{m_0}\left[z_1 \exp\left(-\frac{k_{\perp}^2}{2eB}\right)\right].\label{pi22}\\
\end{equation}
Here $\mu^\prime=(\alpha/2\pi)\mu_B$ is the anomalous magnetic
moment of the electron. The exponential factor in (\ref{pi22})
plays a very important role. If $2 e B\ll k_{\perp}^2 $, it would
make the exponential factor negligible small  and in the limit
$B\to 0$, it vanishes (as well as $\mu_\gamma$ below). In the
opposite case, if $k_{\perp}^2 \ll 2eB$, the exponential is of
order unity. By considering the last assumption and the case of
transversal propagation ($k_{\parallel}=0$),
 the dispersion equation for the second mode has the
solution
\begin{equation}
\omega^2=
k_\parallel^2+k_\perp^2\left(1+\frac{2\mu^{\prime}B}{m_0}\right)^{-1}
\label{de1}
\end{equation}
from which it results that the photon energy can be expressed
approximately as a linear function of the external field B, as
indicated in (\ref{de0}), where
\begin{equation}
\mu_\gamma^{(2)}=\frac{\mu^\prime k_\perp^2}{m_0\omega} \label{de11}
\end{equation}
Notice that, as pointed out above, in the limit $B=0$, if $k_\perp^2
\neq 0$, then $\mu_\gamma^{(2)}$ vanishes. By considering
transversal propagation and $\omega \simeq k_{\perp}$ one can write
\begin{equation}
\mu_\gamma^{(2)}=\frac{\mu^{\prime}\vert\textbf{k}_{\perp}\vert}{
m_0}. \label{lemm}
\end{equation}
As $\mu_\gamma^{(2)}>0$, the magnetic moment is paramagnetic, which
is to be expected since vacuum in a magnetic field behaves as
paramagnetic \cite{Elizabeth}. For photon energies $\omega \sim
10^{-6}m_0$ and $B \sim 10^{4}$G, we have $\mu_{\gamma}\sim
10^{-6}\mu^{\prime}$. In a more exact approximation we must take
into account the contribution from higher Landau quantum numbers.

We will be interested now on the photon magnetic moment in the
region near the thresholds, and for fields $B \lesssim B_c$. The
eigenvalues of the modes can be written approximately \cite{Hugo2}
as
\begin{equation}
\pi_{n,n^{\prime}}^{(i)}\approx-2\pi\phi_{n,n^{\prime}}^{(i)}/\vert\Lambda\vert
\label{eg5}
\end{equation}
with $\vert\Lambda\vert=((k_\perp^{\prime 2 }-k_\perp^{\prime
\prime 2})(k_\perp^{\prime 2}-\omega^2+k_\parallel^2))^{1/2}$
 with $k_\perp^{\prime\prime
2}=m_0^2[(1+2nB/B_c)^{1/2}-(1+2n^{\prime}B/B_c)^{1/2}]^2$, is the
squared threshold energy for excitation between Landau levels
$n,n^{\prime}$ of an electron or positron. The functions
$\phi_{n,n^{\prime}}^{(i)}$ are expressed in terms of Laguerre functions of the variable $k_\perp^2/2e B$.

In the vicinity of the first resonance $n=n^{\prime}=0$ and
considering $k_\perp\neq0$ and $k_\parallel\neq0$, according to
\cite{shabad1,shabad2} the physical eigenwaves are described by
the second and third modes, but only the second mode has a
singular behavior near the threshold and the function
$\phi^{(2)}_{n n'}$ has the structure
\begin{equation}
\phi_{0,0}^{(2)}\simeq-\frac{2\alpha e B m_0^2}{\pi}\textrm{exp}\left(-\frac{k_\perp^2}{2e B}\right)
\end{equation}
In this case
 $k_{\perp}^{\prime\prime 2}=0$ and $k_{\perp}^{\prime
2}=4m_0^2$ is the threshold energy.

By using the approximation given by (\ref{eg5}) the dispersion
equation (\ref{egg})  is turned into a cubic equation in the variable
$z_1$ that can be solved
 by applying the Cardano formula. We
will refer in the following to (\ref{eg2}) as the real solution of
this equation.

We should define the functions $
m_n=(k_\perp^{\prime}+k_\perp^{\prime\prime })/2$, $
m_{n^{\prime}}=(k_\perp^{\prime }-k_\perp^{\prime\prime })/2$ and
$ \Lambda^{*}=4m_nm_{n^\prime} (k_\perp^{\prime 2}-k_\perp^2)$ to
simplify the form of the  solutions (\ref{eg2}) of the equation
({\ref{egg}}). The functions  $f_{i}$ are dependent on
$k_{\perp}^{2},k_{\perp}^{\prime 2},k_{\perp}^{\prime\prime 2},
B$, and are
\begin{equation}
f_i^{(1)}=\frac{1}{3}\left[2k_\perp^{2}+k_\perp^{\prime
2}+\frac{\Lambda^{* 2}}{(k_\perp^{\prime\prime 2}-k_\perp^{\prime
2})\mathcal{G}^{1/3}}+
\frac{\mathcal{G}^{1/3}}{k_\perp^{\prime\prime
2}-k_{\perp}^{\prime 2}}\right] \label{fi}
\end{equation}
where $\mathcal{G} =6 \pi\sqrt{3}D-\Lambda^{*3}+54 \pi^2
\phi_{n,n^{\prime}}^{(i) 2}(k_\perp^{\prime\prime
2}-k_\perp^{\prime 2})^2$ with
\[
D=\sqrt{-(k_\perp^{\prime 2}-k_\perp^{\prime\prime
2})^2\Lambda^{*3}\phi_{n,n^{\prime}}^{(i) 2}\left[1-\frac{27 \pi^2
\phi_{n,n^{\prime}}^{(i)2}(k_\perp^{\prime\prime2}-k_\perp^{\prime
2})^2}{\Lambda^{*3}}\right]}
\]

Besides (\ref{fi}), there are two other solutions of the above-
mentioned cubic equation resulting from the substitution of (\ref{eg5})
in (\ref{egg}). These are complex solutions and are located in the
second sheet of the complex plane of the variable
$z_1=\omega^2-k_\parallel^2$ but they are not interesting to us in
the present context.

Now the   magnetic moment of the photon can be calculated by taking
the implicit derivative $\partial\omega/\partial B$ in the
dispersion equation. From (\ref{egg}) and (\ref{eg5})  it is
obtained that
\begin{widetext}
\begin{equation}
\mu_\gamma^{(i)}=\frac{\pi}{\omega(\vert
\Lambda\vert^3-4\pi\phi_{n,n^\prime}^{(i)}m_n
m_{n^\prime})}\left[\phi_{n,n^{\prime}}^{(i)}\left(A\frac{\partial
m_n}{\partial B}+Q\frac{\partial m_{n^\prime}}{\partial
B}\right)-\Lambda^2\frac{\partial
\phi_{n,n^{\prime}}^{(i)}}{\partial B}\right] \label{mm2}
\end{equation}
\end{widetext} with $
A= 4m_{n^\prime}[z_1+(m_n+m_{n^\prime})(3m_n+m_{n^\prime})] $
and
 $
Q= 4m_{n}[z_1+(m_n+m_{n^\prime})(m_n+3m_{n^\prime})] $.

In the vicinity of the first threshold  $k_\perp^{\prime\prime
2}=0$, $k_\perp^{\prime 2}=4m_0^2$ and $\partial m_n/\partial B=0$
when $n=0$, therefore for the second mode the photon  magnetic moment is given by
\begin{widetext}
\begin{equation}
\mu_\gamma^{(2)}=\frac{\alpha
m_0^3\left(4m_0^2+z_1\right)\exp\left(-\frac{k_{\perp}^2}{2eB}\right)}{\omega
B_c\left[(4m_0^2+z_1)^{3/2}+\alpha m_0^3 \frac{B}{B_c}
\exp\left(-\frac{k_{\perp}^2}{2eB}\right)\right]}\left(1+\frac{k_{\perp}^2}{2eB}\right).\label{FRR1}
\end{equation}
\end{widetext}

\begin{figure}[!htbp]
\includegraphics[width=3in]{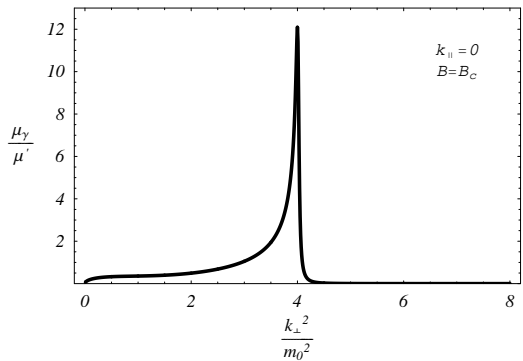}
\caption{\label{fig:Phvk} Photon magnetic moment curve drawn
with regard to perpendicular momentum squared,  for the second
mode $k_\perp^{\prime 2}=4m_0^2$ with $n=n^{\prime}=0$.}
\end{figure}

It is easy to show that this function has a maximum near the
threshold. If we consider $\omega$ near $2m_0$, the function
$\mu_{\gamma}^{(2)}=f(X)$, where $X=\sqrt{4m_0^2-\omega^{2}}$ has a
maximum for $X= {\pi\phi_{00}^{(2)}/m_0}^{1/3}$, which is very close
to the threshold.

Thus, near  the first threshold and in the second mode of propagation
the expression (\ref{mm2}) has a maximum value when
$k_\perp^2 \simeq k_\perp^{\prime 2}$. Therefore in a vicinity of
the first pair creation threshold the magnetic moment of the
photon has a resonance peak which is positive, indicating a paramagnetic behavior, and its value is given by
\begin{equation}
\mu_{\gamma}^{(2)}=\frac{m_0^2(B+2B_c)}{3m_\gamma
B^2}\left[2\alpha\frac{B}{B_c}\exp\left(-\frac{2B_c}{B}\right)\right]^{2/3}
\label{mumax}
\end{equation}
Obviously, (\ref{mumax}) would vanish also for $B\to 0$. The
maximum of (\ref{mumax}) is given numerically by
\begin{equation}
\mu_\gamma^{(2)}\approx 3\mu^\prime
\left(\frac{1}{2\alpha}\right)^{1/3}\approx 12.85\mu^\prime
\end{equation}

Thus, the
maximum value achieved by the photon magnetic moment under the assumed conditions is larger than
twice the anomalous magnetic moment of the electron.

In (\ref{mumax}) we introduced the quantity $m_{\gamma}$ which has meaning near the thresholds, and which could be named as the "dynamical mass"
of the photon in presence of a strong magnetic field, which is defined
by the equation
\begin{equation}
m_{\gamma}^{(2)}=\omega (k_\perp^{\prime
2})=\sqrt{4m_0^2-m_0^2\left[2\alpha
\frac{B}{B_c}\exp\left(-\frac{2B_c}{B}\right)\right]^{2/3}}
\label{dm}
\end{equation}

The "dynamical mass"  accounts for the fact that the massless photon coexists
with the massive pair near the thresholds, leading to a behavior very similar to that
of a neutral massive vector particle bearing a magnetic
moment. However, it does not violate gauge invariance since the condition $\Pi_{\mu \nu}(0,0,B) =0$ is preserved. The idea of a photon mass has been introduced previously, for instance in ref.\cite{Osipov}, in a  regime different from ours, in which  $k_{\parallel}\gg 4m^2$.

\begin{figure}[!htbp]
\includegraphics[width=3in]{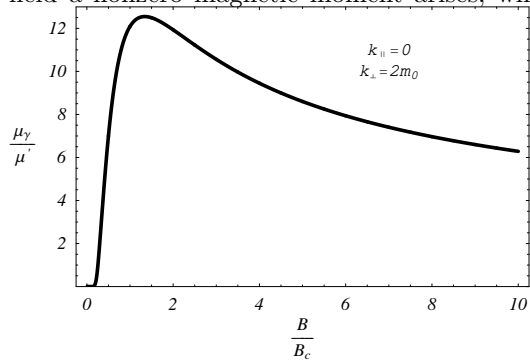}
\caption{\label{fig:mb1} Photon magnetic moment behavior  with
regard to external magnetic field strength for the second mode.}
\end{figure}

We conclude, thus, that for photons in a strong magnetic field a nonzero
magnetic moment arises, which is paramagnetic, and has a maximum near the first threshold of pair creation. These results may have several interesting consequences. For instance, if we consider  a photon beam of density $n_\gamma$, it carries a magnetization ${\cal M}=n_\gamma \mu_{\gamma}^{(2)}$ which contributes to increasing the field $B$ to $B^{\prime}= B + 4\pi{\cal M}$. Trough this mechanism, the radiation field might contribute to the increase of the external field.

Both authors are indebted to A.E. Shabad for several comments and important remarks on the subject of this paper. H.P.R. thanks G. Altarelli, J. Ellis and P. Sikivie for comments, and to CERN, where part of this paper was written,  for hospitality.

\bibliographystyle{apsrev}

\end{document}